\documentclass[aps,prd,showpacs,amsmath,floats,floatfix,twocolumn,nofootinbib]{revtex4}
\usepackage{graphicx,amssymb,amsmath,amsbsy,mathrsfs}
\usepackage{bm}

\usepackage[usenames]{color}
\usepackage{ulem}
\newcommand{\beq}{\begin{equation}}
\newcommand{\eeq}{\end{equation}}
\newcommand{\beqa}{\begin{eqnarray}}
\newcommand{\eeqa}{\end{eqnarray}}

\newcommand{\hmm}{\hspace{1mm}}

\begin{document}
\vspace{-2.5cm} 

\title{The Final Remnant of Binary Black Hole Mergers: Multipolar Analysis}

\author{Robert Owen}

\affiliation{Center for Radiophysics and Space
    Research, Cornell University, Ithaca, NY 14853}

\email{owen@astro.cornell.edu}

\date{September 24, 2009}

\begin{abstract}

Methods are presented to define and compute source multipoles of dynamical 
horizons in numerical relativity codes, extending previous work in the 
isolated and dynamical horizon formalisms to allow for 
horizons that are not axisymmetric.  These methods are then 
applied to a binary black hole merger simulation, providing evidence that 
the final remnant is a Kerr black hole, both through the (spatially) 
gauge-invariant recovery of the geometry of the apparent horizon, and 
through a detailed extraction of quasinormal ringing modes directly from 
the strong-field region.  

\end{abstract}

\maketitle

\section{Introduction}

The problem of the merger of binary black hole systems now seems to be well 
under the control of numerical relativity.  More precisely, the development, 
due to Einstein's vacuum evolution equations, of an initial data set 
containing two apparent horizons into a quiescent state containing only 
one apparent horizon, has now been carried out numerous times by various 
research groups, with somewhat different numerical treatments and mathematical 
formalisms~\cite{Pretorius2005a, Campanelli2006a, Baker2006a, Scheel2008}.  
Numerical relativity is 
now a tool for studying the physics of strong gravitational fields.

When applying this tool, one is immediately faced with a fundamental 
irony of numerical relativity: a numerical code is incapable of dealing with  
abstract tensors, and must instead compute their components in a particular 
vector basis.  
The fundamental physics of general relativity, however, is 
basis independent.  One must be careful to ensure that any physical claims are 
independent (to whatever extent is possible) of the coordinate system and 
vector basis in which they are demonstrated.  

One reasonably well-developed example is the computation of spin angular 
momentum in binary black hole 
simulations.  Numerous investigations have been made of the physics of 
spinning black hole mergers, presenting in some detail effects such as a 
hang-up of the merger, allowing angular momentum to be radiated so 
that the final remnant 
has sub-extremal spin~\cite{Campanelli2006c}; spin 
flips~\cite{campanelli2007b}, in which the dynamics of the merger 
cause the spin direction of the merged black hole to be dominated by the 
direction of orbital angular momentum, rather than the spins of the progenitor 
black holes; and perhaps of most astrophysical interest, 
the kick applied to a merged black hole system, balancing the linear momentum
given off in gravitational radiation during nonsymmetric 
mergers~\cite{Herrmann2007, Koppitz2007, Campanelli2007a, Choi-Kelly-Boggs-etal:2007, Gonzalez2007b, Campanelli2007, Baker2007, Tichy:2007hk, Herrmann2007c, Bruegmann-Gonzalez-Hannam-etal:2007, Schnittman2007, Baker2008, MillerMatzner2008, Healy2008}.
A certain amount of investigation has also gone into the study of black holes 
of nearly-extremal spin in binary configurations, an avenue that could probe 
the limits of cosmic censorship~\cite{DainEtAl:2008, Lovelace2008}.  Because 
such 
physical effects must be parametrized according to the spin angular momenta 
of the dynamical black holes, methods must be devised to define and compute 
such a quantity.  The most common approach begins with a formula that appears 
both in the quasilocal formalism of Brown and York~\cite{BrownYork1993} and in 
the 
isolated and dynamical horizon formalisms~\cite{Ashtekar2001, Ashtekar2003}.  
This formula gives 
angular momentum within a two-surface (normally taken to be an apparent 
horizon of spherical topology) 
as a functional of a vector field tangent to that surface.  This 
vector field is interpreted as a generalized rotation generator, and it is 
through this that the vectorial nature of angular momentum in Newtonian 
mechanics is 
generalized.  In order to apply this formula, a rule must be given for 
choosing such a generalized 
rotation generator on a dynamical black hole.  Methods have recently been 
presented to fix these vector 
fields as ``approximate Killing vectors'' in a precise 
sense~\cite{Dreyer2003, OwenThesis, Cook2007, Lovelace2008, Beetle2008}.

The method presented in Refs.~\cite{Cook2007, Lovelace2008, Beetle2008} 
actually 
provides much more information than just the generalized rotation generators.  
The method starts with the expression of the vector field in terms of a 
scalar potential:
\beq
\phi^A = \epsilon^{AB} \nabla_B z,
\eeq
where uppercase latin letters index the tangent bundle to the two-dimensional 
surface, $\nabla$ is the covariant derivative on this tangent bundle, inherited from that on spacetime, and $\epsilon_{AB}$ is the Levi-Civita tensor on the 
surface.  The vector $\vec \phi$ is said to be an approximate Killing vector 
if it is of this form and if the function $z$ satisfies a certain 
generalized eigenvalue problem on the surface.  On a metric 
sphere\footnote{Throughout this paper, by 
``metric sphere'' we mean a sphere in the metric sense: a closed 2-surface of 
constant positive intrinsic curvature, sometimes also referred to as a 
``round sphere.''}, the operator in this problem reduces to the conventional 
spherical Laplacian, so 
these functions can be interpreted as spherical harmonics of the two-surface.  
In this special case, the three $\ell = 1$ harmonics provide the 
three standard rotation generators.  

The appearance of generalized spherical harmonics in this formalism raises the 
possibility that one could naturally define more than just the spin 
angular momentum (which is often physically understood as the current 
dipole moment of the source).  Perhaps with the help of the 
remaining eigenfunctions, we could define higher multipole moments.  

The idea of quasilocal {\em source multipoles} in general relativity is not 
new.  In Ref.~\cite{AshtekarMultipole2004}, a complete formalism was presented 
for application on axisymmetric isolated horizons.  This formalism involves 
numbers $I_n$ and $L_n$, where $n$ is a nonnegative integer index.  Ashtekar 
et al.~not only provided definitions for these multipole moments, they also 
proved that they completely characterize the isolated horizon geometry, 
that a unique isolated horizon (up to diffeomorphism) can be 
constructed from given multipole moments.  

A few years later, Schnetter, Krishnan, and Beyer~\cite{Schnetter2006} were 
the first to apply this multipole moment formalism in numerically generated 
dynamical spacetimes.  Their work was intended as a wide overview of the use 
of the dynamical horizon formalism in interpreting numerical relativity 
simulations; for them, multipole moments were just one of many points of 
discussion.  They applied the formalism of~\cite{AshtekarMultipole2004} in 
an essentially unmodified form.  Because this construction 
is restricted to axisymmetric horizons, the 
authors of~\cite{Schnetter2006} focused attention on an axisymmetric 
black hole merger.  

Another application of this method appeared in Ref.~\cite{Vasset2009}, a paper 
presenting methods to solve for conformally curved initial data sets.  As one 
might expect, when solving for a fully stationary single-black-hole initial 
data set, the result is a slice of the Kerr spacetime, a fact that the 
authors confirm using the multipole construction 
of Ref.~\cite{AshtekarMultipole2004}.

Quite recently, another paper 
appeared~\cite{Jasiulek2009} which introduced a novel 
scheme for computing multipole moments indirectly, from surface integrals of 
various powers of the curvature.  This new method is still restricted to 
axisymmetric horizons, but it avoids the need to explicitly find the 
axisymmetry, and could markedly improve accuracy in cases where it can be 
used. 

Here, we take a slightly different approach.  Rather than directly applying 
the 
methods of Ref.~\cite{AshtekarMultipole2004} in an axisymmetric merger, we modify 
the method, in a manner briefly suggested by its authors, so that it can be 
applied without the requirement of axisymmetry.  Whereas the original method 
in Ref.~\cite{AshtekarMultipole2004} involved a preferred coordinate system on the 
axisymmetric horizon, in which spherical harmonic projections could be taken, 
we choose to 
project the relevant quantities against spectrally-defined spherical 
harmonics.  Such harmonics are invariantly defined on any given topological 
sphere endowed with intrinsic geometry, as eigenfunctions of geometric 
operators, such as the one mentioned above relevant to the computation 
of spin angular momentum.  Extra structure, such as 
the axisymmetry that provides the preferred coordinate system 
of Ref.~\cite{AshtekarMultipole2004}, is not necessary.  While the continuum 
eigenvalue problems that define these harmonics  
would complicate analytical treatments, they are quite straightforward to 
solve numerically.  

In section~\ref{s:SphHarms} we introduce the details of this method, 
in particular the eigenvalue problems used to define spherical harmonics 
on deformed spheres.  In section~\ref{s:Results}, 
we investigate one of the simplest applications of current physical 
relevance.  This is the question of the final remnant of a numerical merger 
of two vacuum black holes.  While the general expectation is that the remnant 
of such mergers will generically be a Kerr black hole, relatively little 
effort 
has gone into a detailed investigation of whether this is actually the case.  
This question is of relevance to the status of black hole uniqueness, whose 
rigorous proof still involves certain analyticity 
assumptions~\cite{FriedrichRaczWald1999}.  It is also related to the question 
of stability of the Kerr solution, which has so far 
been proven only for individual modes of linear 
perturbations~\cite{Whiting1989}.  Even if we fully accept the expectation 
that general relativity 
must force the remnant of a black hole merger to be Kerr, the detailed 
recovery of the Kerr solution at late times, in as gauge-invariant a manner as 
possible, provides at the very least a stringent and physically-relevant 
code test.  In Ref.~\cite{Campanelli2008}, Campanelli et al.~demonstrated 
that a particular black hole merger simulation approaches Petrov type D in a 
certain 
sense at late times, and carries no NUT charge.  This fact largely 
confirms that their merger produces a Kerr geometry.  One 
advantage of their approach is that it is fully local, that one can 
investigate the approach to Kerr geometry throughout the spatial slices, 
rather than simply on the horizon as we do here.  In a followup to 
the current paper, we intend to repeat many of the methods 
of Ref.~\cite{Campanelli2008} on the datasets discussed in Sec.~\ref{s:Results}.  
Here we focus on multipole moments partly as a complementary method of black 
hole characterization, but also because these moments 
are of interest in their own right, as tools for probing the 
physics of tidal structure in strong-field gravity.

\section{Generalized Spherical Harmonics}
\label{s:SphHarms}

The definitions given in Ref.~\cite{AshtekarMultipole2004} for the mass and current 
multipoles on isolated horizons are very simple spherical harmonic projections 
of quantities related to the intrinsic and extrinsic geometry of the 
apparent horizon\footnote{In the case of isolated horizons, the surfaces of 
interest are arbitrary 
two-dimensional spacelike slices of the three-dimensional null 
isolated horizon.  In the case of dynamical horizons, the two-surfaces of 
interest are the apparent horizons into which the dynamical 
horizon is naturally foliated.} in 
spacetime.  
\beqa
I_\alpha := \oint y_\alpha R \hmm dA,\label{e:MassMoment}\\
L_\alpha := \oint y^B_\alpha \omega_B \hmm dA.\label{e:CurrentMoment}
\eeqa
Here, $dA$ refers to the metric volume element on the apparent horizon, $R$ is 
its intrinsic scalar curvature (not to be confused with 
the Ricci scalar of the full spacetime, or of the spatial slice, or of the 
horizon worldtube), and 
$\omega_A$ is a connection on the normal bundle of the two-surface, which is 
conveniently written in terms of the two future-directed null normals, 
$\vec \ell$ and $\vec n$:
\beq
\omega_A := e^\mu_A \hmm n_\nu \hmm {}^{(4)}\nabla_\mu \ell^\nu,\label{e:omega}
\eeq
where ${}^{(4)}\nabla$ is the metric-compatible torsion-free spacetime 
covariant derivative, and $\{\vec e_A\}$ are basis vectors tangent to the 
two-surface.  Throughout this paper, capital latin letters will index this 
two-dimensional tangent bundle.  The null normals $\vec \ell$ and $\vec n$ 
are, as usual, normalized such that $\vec \ell \cdot \vec n = - 1$.  
In most numerical papers and codes, $\omega_A$ is written and computed in 
terms of the extrinsic curvature of the spatial slice.  Here, we will refer 
to the $I_\alpha$ as the {\em mass multipoles} and the $L_\alpha$ as the 
{\em current multipoles}, though as noted in Ref.~\cite{AshtekarMultipole2004} 
extra factors involving horizon areas and quasilocal spins must be included 
if one wishes to make them dimensionally consistent with the standard 
definitions of these quantities.

The objects $y_\alpha$ appearing in \eqref{e:MassMoment} and $y^A_\alpha$ 
appearing in \eqref{e:CurrentMoment} are scalar and vector spherical 
harmonics, respectively.  It is in the definition of 
these harmonics that the breaking of axisymmetry has the most 
immediately-apparent cost, and therefore where the work in this paper will 
depart most strongly from the construction in Ref.~\cite{AshtekarMultipole2004}.  
In that paper, attention is focused on the case of axisymmetric isolated 
horizons.  Axisymmetry provides a natural coordinate system on the 
apparent horizon, so the spherical harmonics used 
in Ref.~\cite{AshtekarMultipole2004} are the standard ($m = 0$) ones of spherical 
coordinates, applied in this canonical coordinate system.  In other words, 
they are eigenfunctions not of the geometric Laplacian on the apparent 
horizon, but rather of the Laplacian of a metric sphere in these coordinates.  
There is nothing inherently wrong with such a choice in axisymmetry, in fact 
it provides certain benefits in that context\footnote{One such benefit is that 
the mass dipole moment always turns out to be zero.  In 
other words, their construction guarantees that one is in a ``center of mass 
frame.''  In general, this may not hold in our construction, though we have 
not yet seen an example where it fails.}, but 
for purposes of strongly dynamical, 
strongly non-axisymmetric systems a more general approach is called for.  

\subsection{Scalar Spherical Harmonics}
\label{s:ScalSphHarm}

Our approach will be to define the spherical harmonics spectrally, as 
eigenfunctions of the geometric Laplacian operator (or certain generalizations 
thereof) on the apparent horizon surface.  In other words, our scalar 
spherical harmonics are taken to be the functions $y_\alpha$ that satisfy 
the equation
\beq
\Delta y_\alpha = \lambda_{(\alpha)} y_\alpha\label{e:ssheigprob}
\eeq
for some constant, $\lambda_{(\alpha)}$.  
The function $y_\alpha$ is defined only on the apparent horizon, and $\Delta$ 
is the intrinsic Laplacian of the apparent horizon, 
$\Delta := g^{AB} \nabla_A \nabla_B$.  The letter $\alpha$ 
is a label for the various solutions to the eigenproblem.  

Because the Laplacian in \eqref{e:ssheigprob} 
reduces to the standard spherical Laplacian when the surface becomes a 
metric sphere, the functions $y_\alpha$ reduce to the standard spherical 
harmonics in that special case as well.  However, this is not the only 
self-adjoint operator with this property.  For example, we can 
consider the problem:
\beq
\Delta y_\alpha + q R y_\alpha = \lambda_{(\alpha)} y_\alpha,
\eeq
where $R$ is again the intrinsic scalar curvature of the surface and $q$ is 
a numerical parameter.  In the case of a metric sphere, where $R$ is constant, 
the second term on the left side does not alter the eigenfunctions, 
it merely increases each eigenvalue.  This eigenproblem, therefore, can again 
be considered to define a reasonable generalization of coordinate spherical 
harmonics.  However, on a deformed sphere, where $R$ is not constant, these 
generalized spherical harmonics will no longer agree with those defined 
by~\eqref{e:ssheigprob}.  To fix this arbitrariness, and since we see no 
particular reason to prefer any other value for $q$, we choose $q=0$, in 
other words the problem in Eq.~\eqref{e:ssheigprob}, to define our scalar 
spherical harmonics.  In the case of vector spherical harmonics, we will see a 
geometrical reason to prefer a particular value for an analogous parameter.  

\subsection{Vector Spherical Harmonics}
\label{s:VecSphHarm}

We will take our generalized vector spherical harmonics to be tangent to the 
surface, in which case they can be written in terms of gradients of two scalar 
potentials:
\beq
y^A_\alpha = \nabla^A w_\alpha + \epsilon^{AB} \nabla_B z_\alpha.
\eeq
Here $\nabla$ is the torsion-free metric-compatible derivative on the apparent 
horizon, and $\epsilon_{AB}$ is the Levi-Civita tensor on it.  
To consider the importance of these two 
potentials we should investigate the one-form $\omega_A$ against which the 
vector spherical harmonics will be projected.  
In Eq.~\eqref{e:omega}, the future-directed null 
vectors $\vec \ell$ and $\vec n$ are orthogonal to 
the apparent horizon and normalized relative to one another by 
the standard Newman-Penrose condition $\vec n \cdot \vec \ell = -1$, but are 
otherwise free.  One can arbitrarily scale the $\vec \ell$ 
vector at the cost of inversely scaling the $\vec n$ vector.  This ``boost 
freedom'' is a standard gauge degree of freedom in the dynamical horizon 
formalism.  The dynamical horizon worldtube carries with it a preferred 
slicing into apparent horizons, but this slicing is only of the dynamical 
horizon itself.  There is no preferred way of extending this slicing into the 
ambient spacetime.  If we wish for our horizon multipoles to be independent 
of this gauge freedom, then we must choose harmonics that project out only 
the gauge invariant part of $\omega_A$.  

From Eq.~\eqref{e:omega}, it is apparent that a boost, 
$\vec \ell \mapsto a \vec \ell$, $\vec n \mapsto a^{-1} \vec n$, will 
add a pure gradient to $\omega_A$:
\beq
\omega_A \mapsto \omega_A - \nabla_A \log(a).
\eeq
Any part of $\omega_A$ that is a pure gradient is therefore entirely due to 
boost gauge, in the sense that it can be transformed away by an appropriate 
boost.  Vector spherical harmonics of the form 
$y^A_\alpha = \nabla^A w_\alpha$ will 
pick up this gauge-dependent information in the 
integral~\eqref{e:CurrentMoment}, however vector spherical harmonics of the 
form $y^A_\alpha = \epsilon^{AB} \nabla_B z_\alpha$ will not.  We therefore 
restrict all 
attention to vector spherical harmonics of this latter form.  

We now need a rule to define the potential functions $z_\alpha$ that appear in 
these 
vector spherical harmonics.  In the case of a metric sphere, the obvious 
choice is that they be the scalar spherical harmonics.  As in the previous 
subsection, there are many ways to generalize the spherical harmonics of 
the metric sphere.  For the current purposes, there is reason to prefer a 
somewhat complicated fourth-order generalized eigenproblem:
\beq
\Delta^2 z_\alpha + \nabla^A \left( R \nabla_A z_\alpha \right) = \lambda_{(\alpha)} \Delta z_\alpha. \label{e:GenEigProb}
\eeq
This generalized eigenproblem also defines the potentials for the approximate 
Killing vector fields used for computing spin angular momentum 
in Ref.~\cite{Lovelace2008}.  For this reason, when this problem is used to define 
the vector spherical harmonics, the current dipole moment of the horizon is 
identical to the quasilocal spin defined there, a quantity that itself reduces 
on axisymmetric isolated horizons to the quasilocal spin defined by 
Hamiltonian 
methods~\cite{Ashtekar2001}.  In Ref.~\cite{AshtekarMultipole2004}, the agreement 
of the current dipole with the 
spin is cited as a reason to prefer using coordinate harmonics in a canonical 
coordinate system rather than spectrally-defined harmonics.  There it was 
assumed that such harmonics would be simple eigenfunctions of the Laplacian, 
like the scalar spherical harmonics of the previous subsection, in which 
case the current dipole would not agree with the standard spin angular 
momentum.  We have averted this situation simply by choosing a better 
operator.  

We should also note that when vector spherical harmonics are chosen in this 
way, we are assured that there will be no current monopole moment.  This 
fact can be viewed in a number of related ways.  On the simplest level, there 
is the fact that when the vector spherical harmonics are defined to be of the 
form $y_\alpha^A = \epsilon^{AB} \nabla_B z_\alpha$, then a potential of 
the form $z_\alpha = {\rm const.}$ cannot define a (normalizable) vector 
spherical 
harmonic.  In some sense, $z_\alpha = {\rm const.}$ can be viewed as a 
solution to Eq.~\eqref{e:GenEigProb} 
with arbitrary eigenvalue, but it is not a well-behaved solution.  The 
generalized eigenproblem is technically {\em singular} in function 
spaces that include constants~\cite{LapackUsersGuide}, meaning that well-behaved solutions cannot be found unless the 
function space is restricted to, for example, functions with zero average 
over the sphere, a condition which removes all nonzero constants from 
consideration.  

Another way of looking at this, which helps to elucidate the relationship 
between the mass and current multipoles, is that when the vector spherical 
harmonics are defined in this way, an integration by parts allows the current 
moments to be written as: 
\beq
L_\alpha := \oint z_\alpha \Omega \hmm dA, \label{e:MomentsFromSpinFn}
\eeq
where $\Omega := \epsilon^{AB} \nabla_A \omega_B$ can be interpreted 
geometrically as a scalar curvature of the normal bundle of the 
two-dimensional surface in four-dimensional spacetime.  The current moments 
thus represent for the extrinsic geometry of the apparent horizon what the 
mass moments represent for its intrinsic geometry.  

The complex combination of these two curvatures, $R + i \Omega$, is sometimes 
called the {\em complex curvature} of the two-surface embedding.  As is 
briefly described in Sec.~4.14 
of Ref.~\cite{PenroseRindler}, the vanishing of the current monopole moment 
can be understood geometrically in this context as a result of the 
generalization of the 
Gauss-Bonnet theorem to the lorentzian normal bundle.  The integral of any 
constant multiple of $\Omega$ is a topological invariant, just like that of 
$R$, but because the gauge group on the normal bundle is topologically 
trivial, this invariant must always vanish.

One final point to note, with regard to both the scalar and the vector 
harmonics, is that of normalization.  Solutions of the eigenproblems in 
Eqs.~\eqref{e:ssheigprob} and~\eqref{e:GenEigProb} are determined 
only up to constant multiplicative factors\footnote{In fact, solutions of 
Eq.~\eqref{e:GenEigProb} are determined only up to constant multiplicative 
and additive factors, however additive constants have no effect on the 
multipoles due to the vanishing of the current monopole.}.  We fix these 
factors with an integral normalization condition.  The condition 
imposed on scalar spherical harmonics is:
\beq
\oint \left( y_\alpha \right)^2 \hmm dA = 1. \label{e:sshnorm}
\eeq
On metric spheres in 
Euclidean space, this reduces to the standard normalization condition for 
scalar spherical harmonics (up to a factor of areal radius).  

For vector spherical harmonics, the normalization condition we impose is:
\beq
\oint g_{AB} y^A_\alpha y^B_\alpha \hmm dA = 1. \label{e:vshnorm}
\eeq
This differs slightly from the standard normalization condition for axial 
vector spherical harmonics in euclidean space, which involves an extra factor 
of $\ell (\ell + 1)$, but because the generalization of the index $\ell$ is 
not an integer, but rather a function of the eigenvalue $\lambda_{(\alpha)}$, 
we simply leave this factor out.  

Normalization conditions like those above still don't determine a sign for the 
spherical harmonics.  This sign ambiguity translates directly into a sign 
ambiguity for the multipoles.  We fix the sign with the condition that the 
values of the multipole moments be nonnegative.  

In summary, the approach we take to defining multipoles in numerical 
simulations begins with finding solutions $y_\alpha$ and $z_\alpha$ of the 
eigenproblems \eqref{e:ssheigprob} and \eqref{e:GenEigProb} defined on the 
apparent horizon two-surface.  From the 
potential $z_\alpha$, vector spherical harmonics are computed as 
$y_\alpha^A := \epsilon^{AB} \nabla_B z_\alpha$.  The harmonics are 
normalized with the conditions in Eqns.~\eqref{e:sshnorm} 
and~\eqref{e:vshnorm}.  Then, the surface integrals in 
Eqns.~\eqref{e:MassMoment} and~\eqref{e:CurrentMoment} are computed to be 
the multipoles.

\section{Numerical Results}
\label{s:Results}

The immediate purpose of this mathematical machinery is to investigate the 
remnant of a binary black hole merger.  There is a very large space of 
physically-relevant mergers worth investigating, including variations in the 
initial mass ratio, eccentricity, spin magnitudes, and spin directions.  
For this paper we will focus on a very simple case: the merger of a 
non-eccentric binary of equal mass, nonspinning black holes.  This data set 
is discussed in detail in Ref.~\cite{Scheel2008}, which briefly 
notes the fact that two independent measures of the final spin agree to 
well within their expected numerical errors.  This claim can be considered a 
first indication that the tidal structure of the quiescent black hole is 
that of Kerr, as this is the case in which these two measures of spin are 
designed to agree.  Our goal now is to present the rest of the tidal 
information, to the extent that it can be resolved in the code, to strengthen 
the case that the final remnant is a Kerr black hole.  

The code used to compute these multipoles is a part of the Spectral Einstein 
Code ({\tt SpEC}) developed and maintained by the Caltech and Cornell 
Numerical Relativity 
groups, particularly Lawrence E.~Kidder, Harald P.~Pfeiffer, and Mark 
A.~Scheel.  Once an apparent horizon has been found, using 
the method described in Ref.~\cite{Gundlach1998}, the code interpolates all 
relevant data to a pseudospectral grid on that surface.  Because this 
grid is pseudospectral, the code can automatically transform any smooth 
function on the apparent horizon into a truncated expansion in coordinate 
spherical harmonics.  This expansion, inserted into Eq.~\eqref{e:ssheigprob} 
or~\eqref{e:GenEigProb}, provides a finite-dimensional matrix eigenproblem 
(or generalized eigenproblem in the latter case) which is solved using 
the {\tt LAPACK} routine {\tt dggev}.  We emphasize that while this 
construction involves coordinate spherical harmonics, they are only used 
to supply the numerical discretization, so they disappear in the 
continuum limit.  The {\em geometrical} spherical harmonics $y_\alpha$ and 
$y_\alpha^A$ that define the multipoles are, apart from numerical truncation 
error, uniquely defined on any given two-sphere (up to possible degeneracy 
in the eigenspaces).  

The information that we can assess includes not only the values of the 
multipole 
moments defined in Eqns.~\eqref{e:MassMoment} and~\eqref{e:CurrentMoment}, but 
also the spectrum of eigenvalues in 
Eqns.~\eqref{e:ssheigprob} and~\eqref{e:GenEigProb}.  A particular motivation 
for investigating eigenvalues of geometric operators is that they provide 
an indication of symmetries in the horizon.  As is familiar from 
elementary 
quantum mechanics, a symmetry in an operator leads to degeneracies in its 
eigenspaces.  The converse is not necessarily true, but on an intuitive level 
we may interpret 
degeneracies in the eigenspectrum as indicators of possible symmetry.  

\begin{figure}
\includegraphics[scale=0.35]{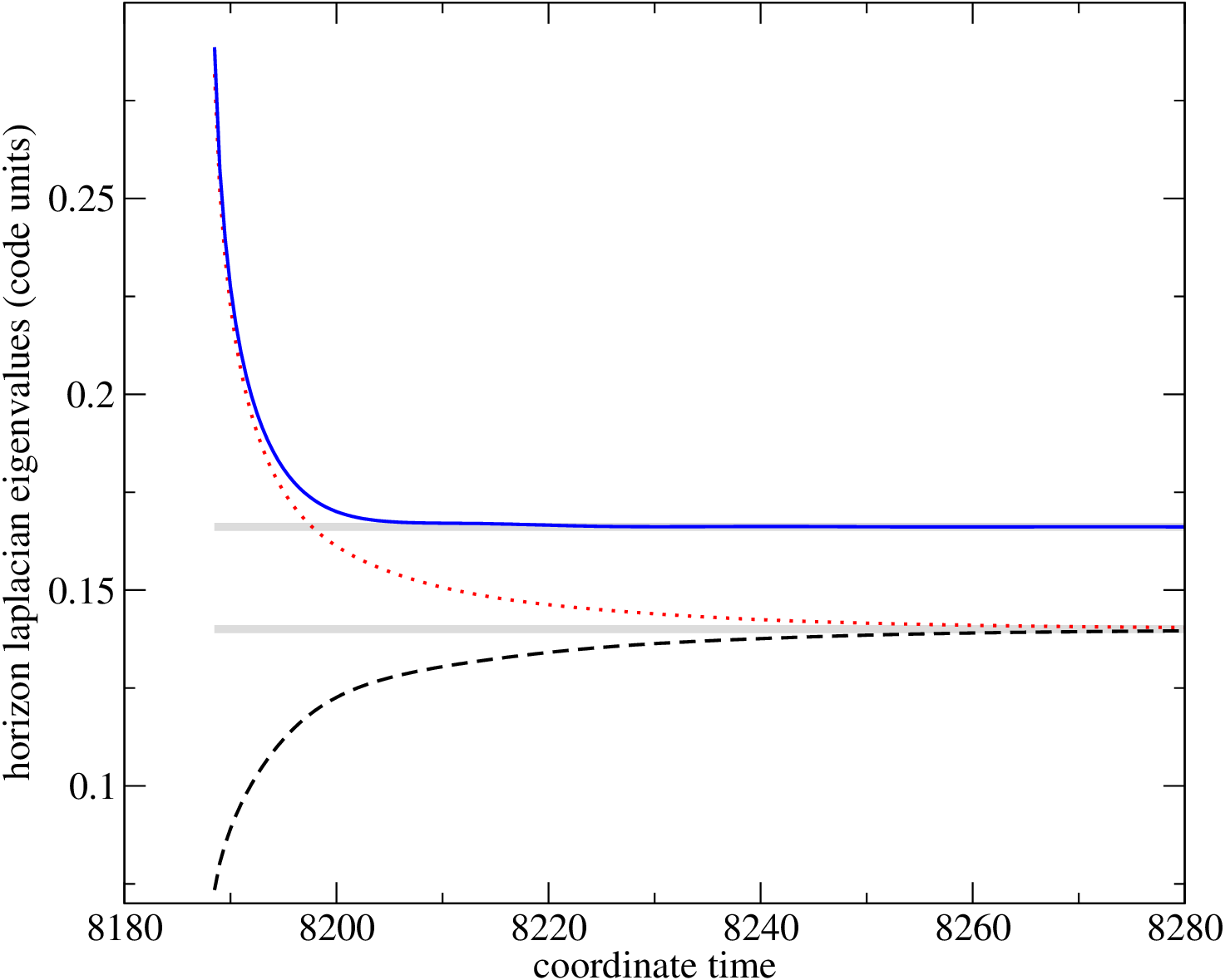}
\caption{ \label{f:LaplacianDipoleEigs} Absolute values of the lowest three 
(dipole) nontrivial 
eigenvalues of the Laplacian on the dynamical horizon during 
the ringdown.  The breaking away of the red (dotted) curve from the 
blue (solid) curve at early 
times is due to the breaking of the initial (approximate) ``peanut'' 
axisymmetry.  The joining of this curve onto the black (dashed) curve at 
late times is 
due to the late-term axisymmetry of the final Kerr 
horizon.  The horizontal gray lines represesent the expected eigenvalues on a 
Kerr horizon with mass and spin equal to the final measured values in the 
simulation.  Thus the convergence of the eigenvalues to these lines 
demonstrates the approach to Kerr geometry.}
\end{figure}

\begin{figure}
\includegraphics[scale=0.35]{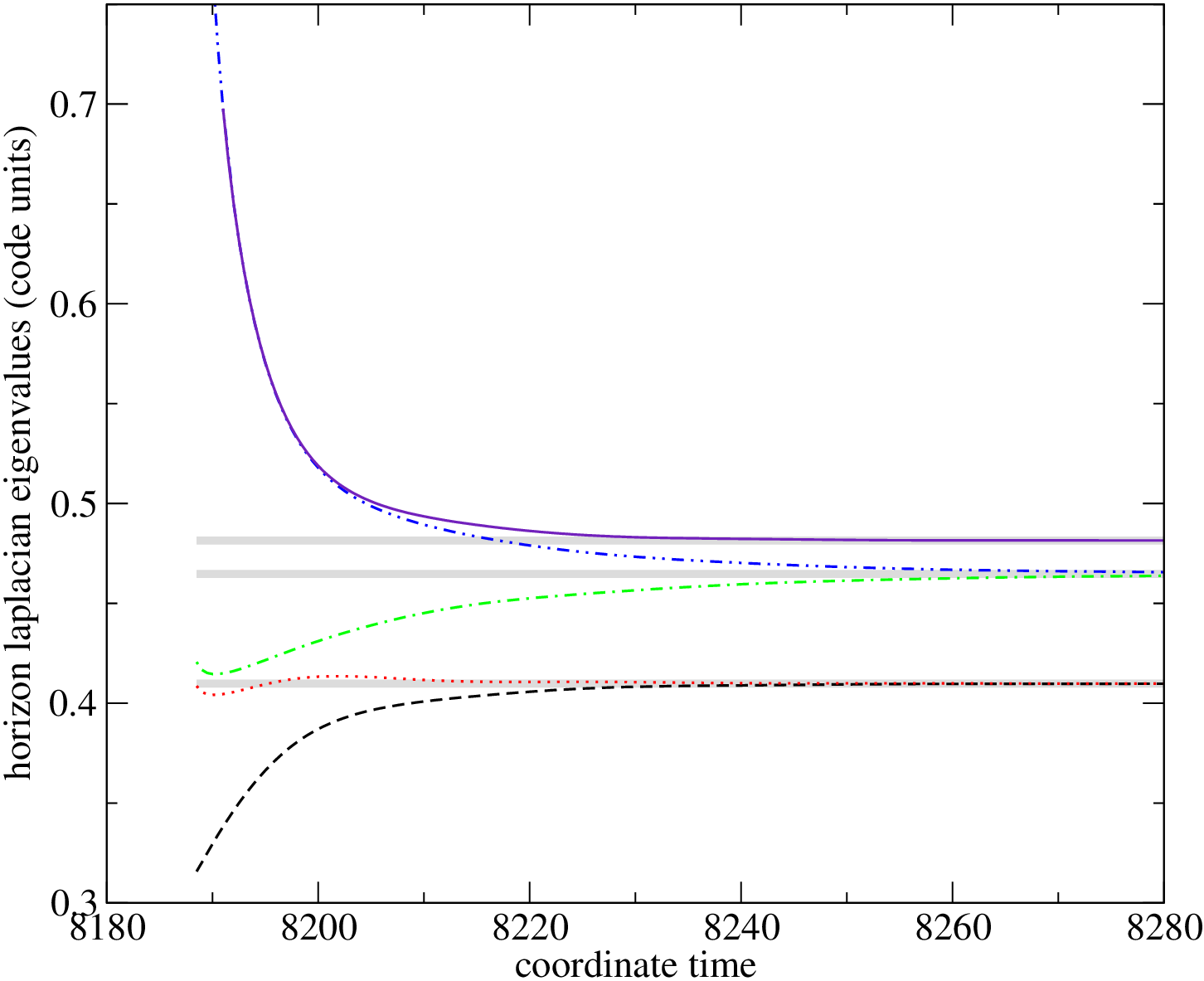}
\caption{ \label{f:LaplacianQuadrupoleEigs} Absolute values of the next five 
(quadrupole)
eigenvalues of the Laplacian on the dynamical horizon during the ringdown.  
As in Fig.~\ref{f:LaplacianDipoleEigs}, the curves are paired 
up at early times, split their degeneracies, and connect in a different 
pairing at late times, again indicating transition from one axis of symmetry 
to another.  The fact that one of these degenerate pairs at early times 
is visibly nondegenerate indicates imperfection in the ``peanut'' axisymmetry 
intuitively due to phase offset tidal bulges built up during the inspiral. 
}
\end{figure}

\begin{figure}
\includegraphics[scale=0.35]{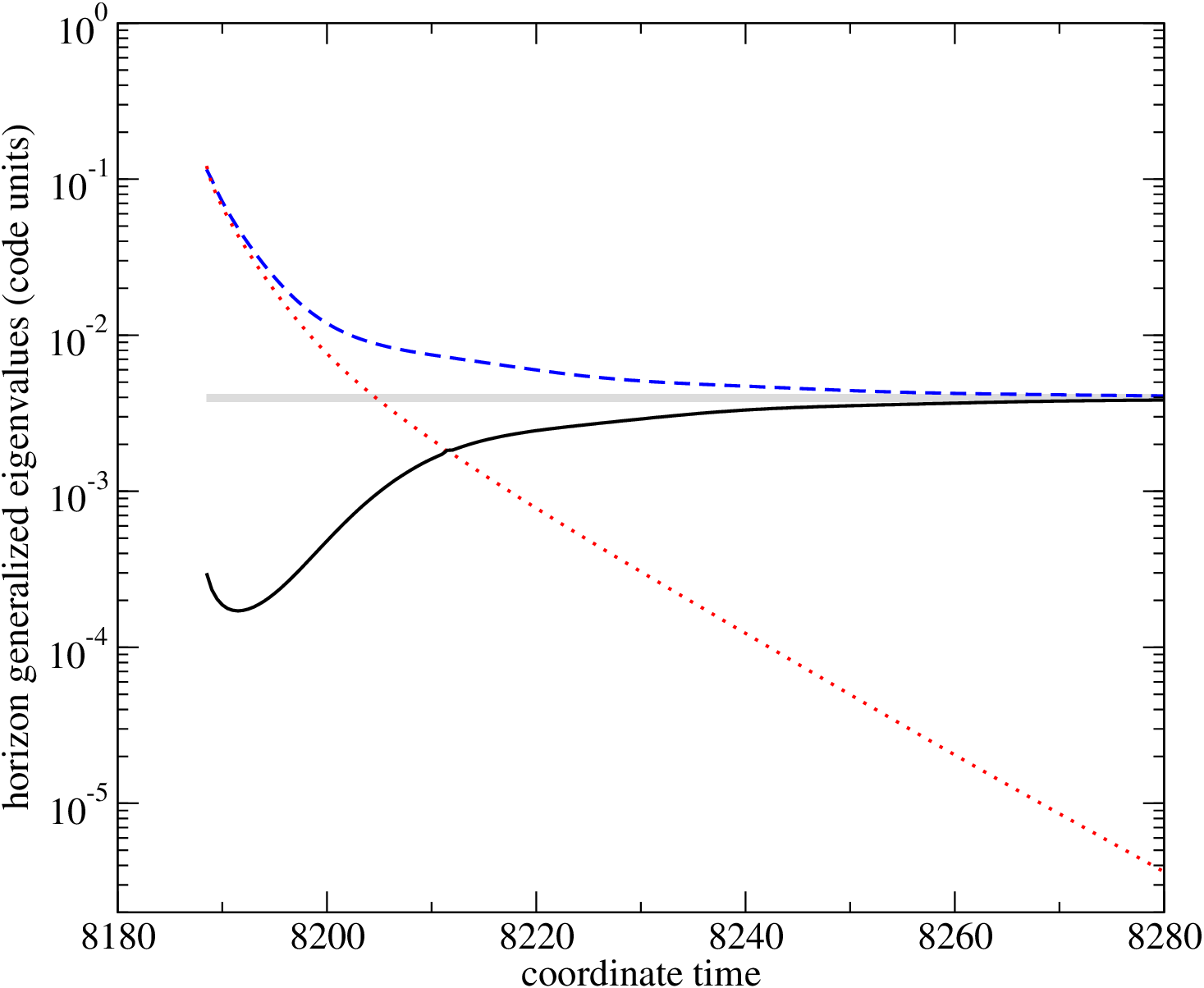}
\caption{ \label{f:AKVOpDipoleEigs} Absolute values of the lowest three 
(dipole) nontrivial 
eigenvalues of the generalized eigenvalue problem in Eq.~\eqref{e:GenEigProb} 
on the dynamical horizon during 
the ringdown.  The vertical axis is now scaled logarithmically to better show 
the approach of the smallest eigenvalue to zero.  As argued in Appendix A 
of Ref.~\cite{Lovelace2008}, the vanishing of this smallest eigenvalue is 
direct evidence of a rotational symmetry of the intrinsic surface geometry, so 
this figure provides a clear picture not only of the symmetry transition 
itself, but also of the relationship between symmetries and degeneracies.  
In particular, the crossing of the red and black (dotted and solid) curves 
can be seen as an 
example of ``accidental'' degeneracy, degeneracy that is not necessitated by 
a symmetry of the operators.}
\end{figure}

\begin{figure}
\includegraphics[scale=0.35]{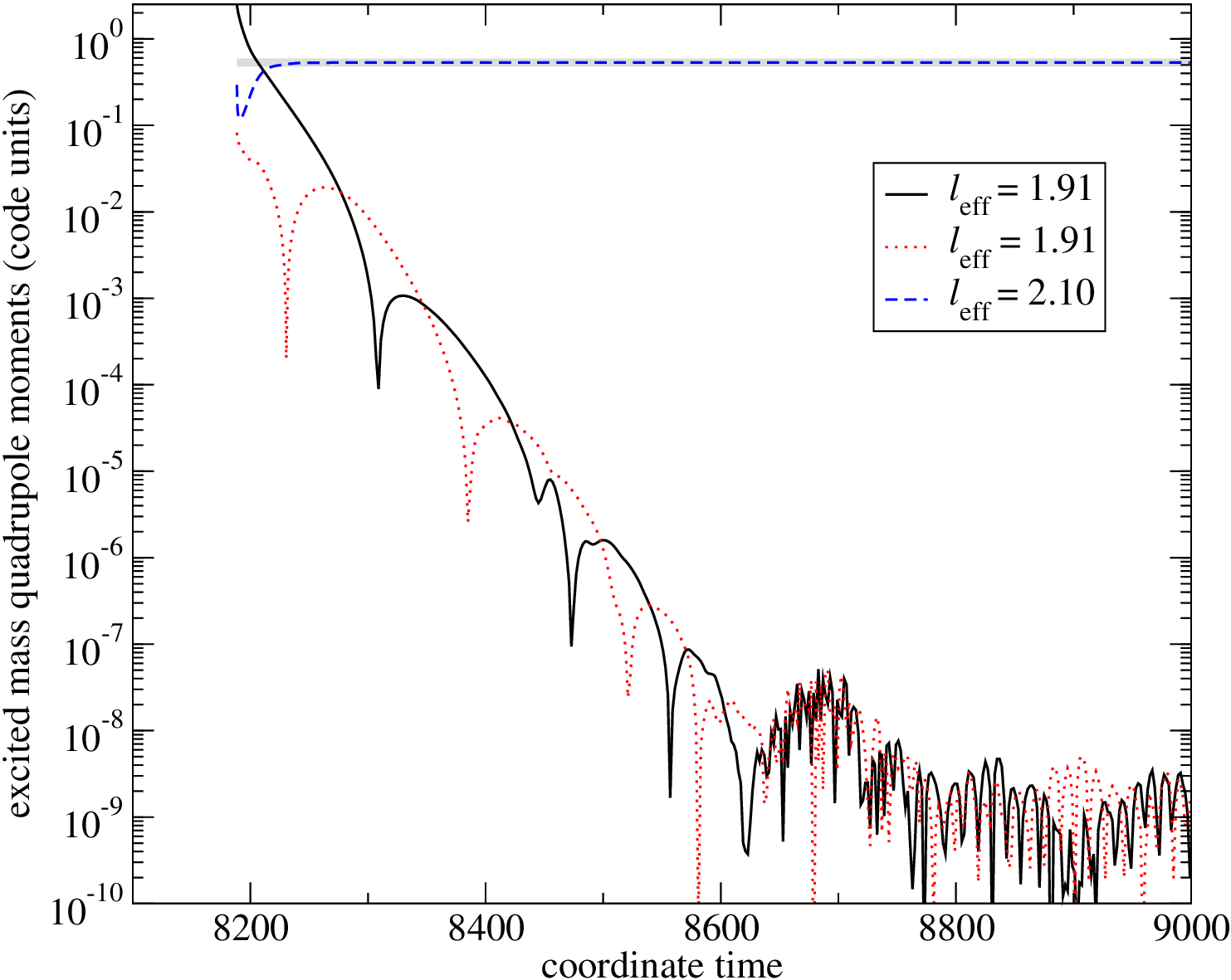}
\caption{ \label{f:quadmoments} Relaxation of the three excited quadrupole 
moments to expected Kerr values.  Two of the five possible quadrupole moments 
must vanish due to the reflection symmetry in the problem, and indeed, their 
computed values are small enough to be considered zero to within ordinary 
numerical 
errors.  Of the remaining three multipoles, two fall exponentially toward 
the level of numerical truncation, and the third quickly settles to the 
expected value for a Kerr black hole of the same final mass and spin.  This 
expected value is shown in the thick horizontal gray line, which for most of 
the simulation overlaps the dashed blue curve.  In this and later figures, 
curves are labeled by the ``effective'' $\ell$ values of their corresponding 
spherical harmonic at late times.  That is, $\ell_{\rm eff}$ is defined from 
the value of the eigenvalue $\lambda$ at late times by the 
equation $\lambda = -\ell_{\rm eff} (\ell_{\rm eff}+1)/r^2$, where $r$ is the 
areal radius of the horizon, again measured after the hole settles down.
}
\end{figure}

\begin{figure}
\includegraphics[scale=0.35]{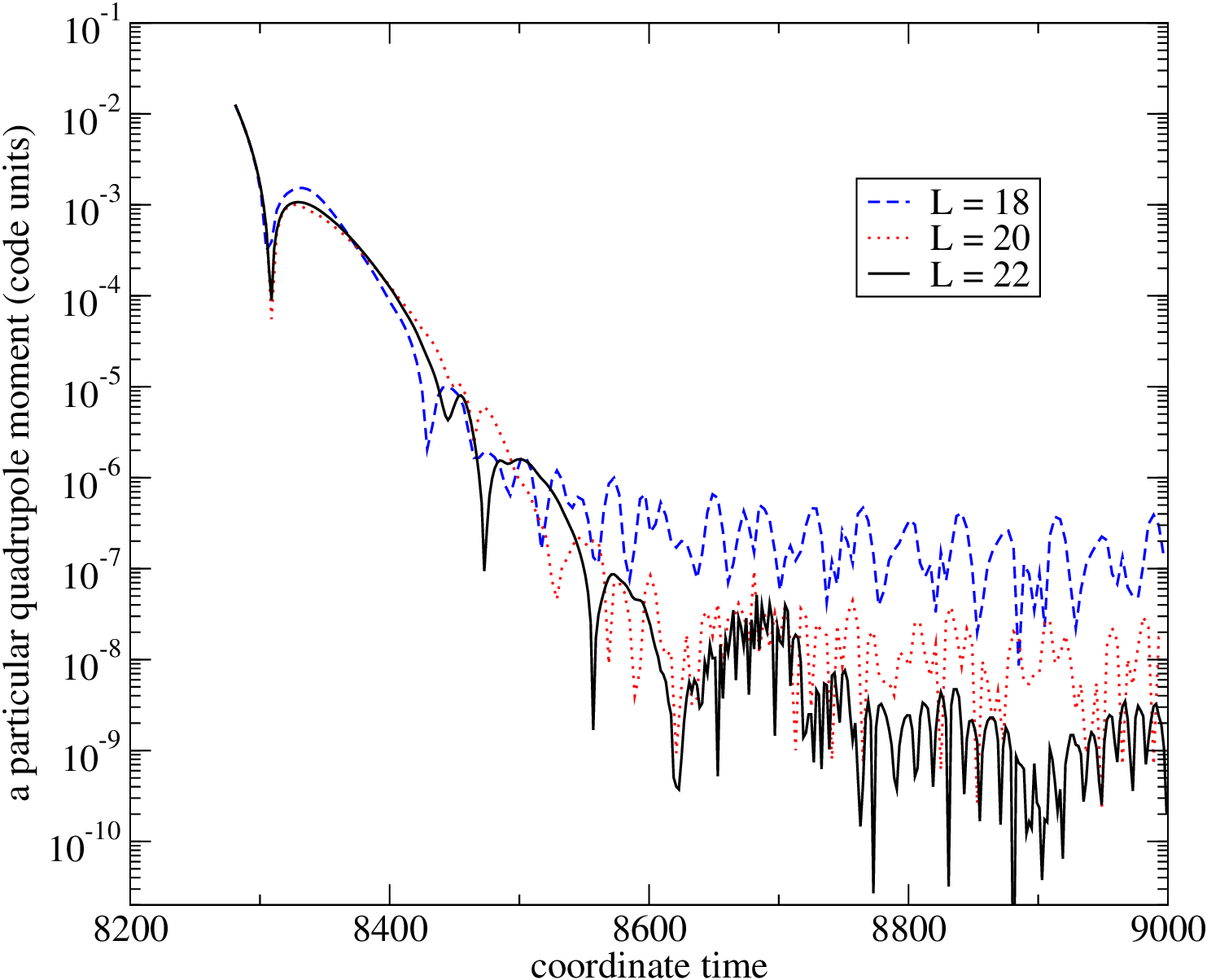}
\caption{ \label{f:quadconvergence}  A particular quadrupole moment from 
Fig.~\ref{f:quadmoments} shown for three different resolutions.  The order of 
the pseudospectral angular discretization is given by $L = 18, 20, 22$ 
respectively, $L$ representing the maximum $\ell$-value of coordinate 
spherical harmonics used to discretize the problem.  At late times, the 
exponential falloff halts, but the level where this occurs converges 
exponentially toward zero as $L$ is increased.  These nonzero values can 
therefore be attributed to standard truncation error.  
}
\end{figure}

\begin{figure}
\includegraphics[scale=0.35]{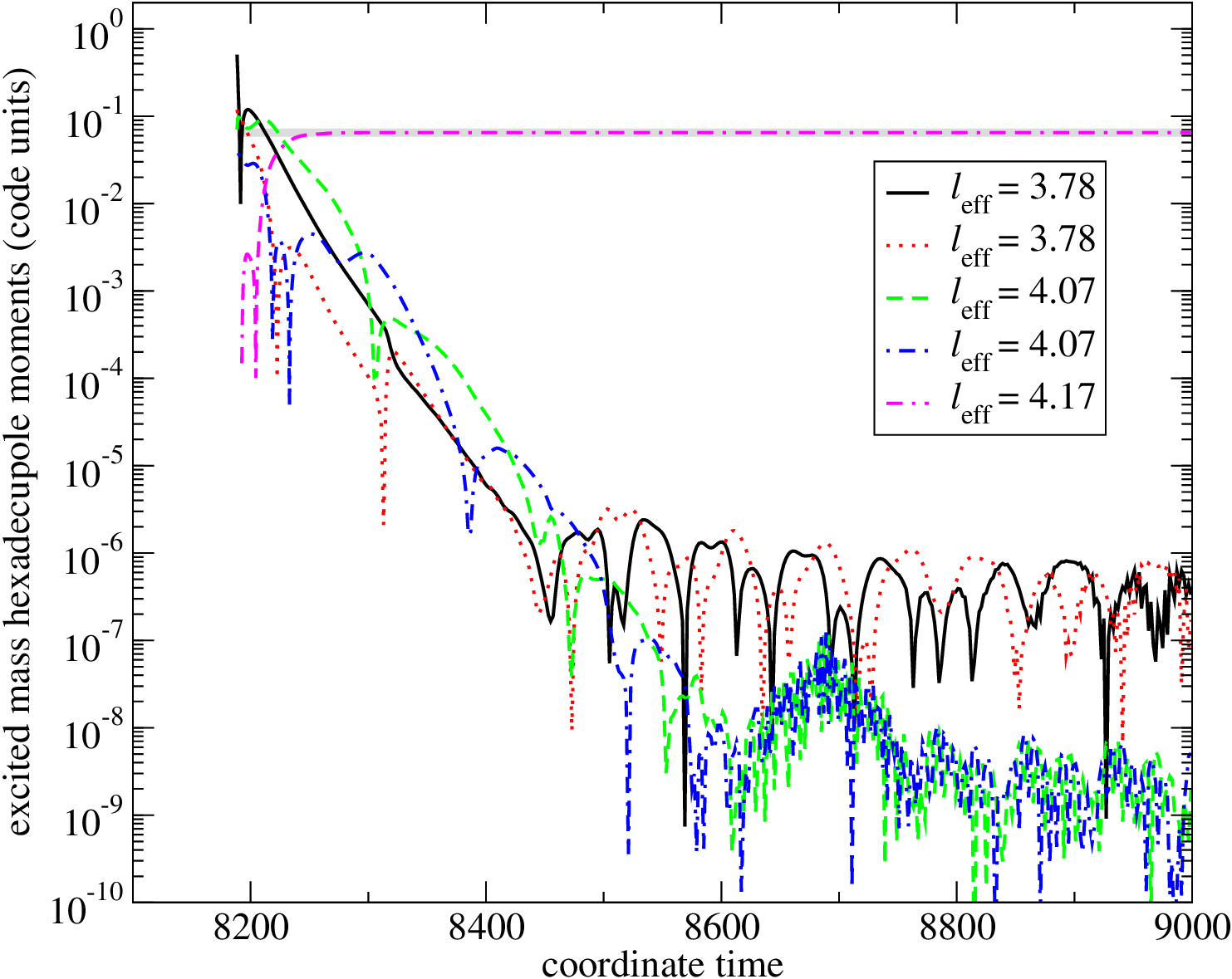}
\caption{ \label{f:hexadecmoments} Relaxation of the five excited hexadecupole 
mass moments to expected Kerr values.  Four of the nine possible moments 
vanish due to the reflection symmetry in the problem.  Of the remaining five, 
four fall to zero exponentially, and the other rises to the expected value for 
a Kerr black hole.
}
\end{figure}

\begin{figure}
\includegraphics[scale=0.35]{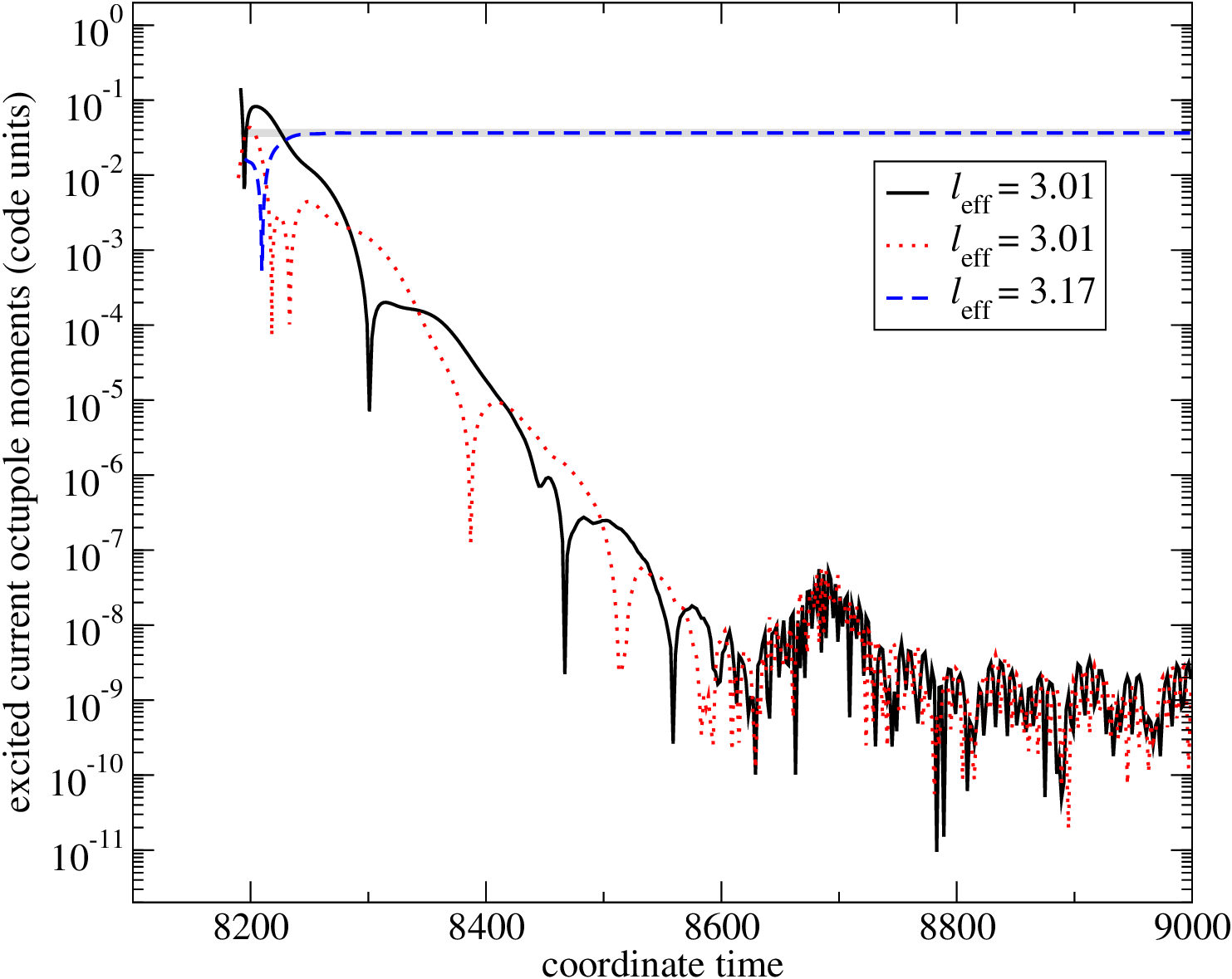}
\caption{ \label{f:magoctupole} Of the seven current octupole moments, only 
three are 
allowed by the reflection symmetry to be excited.  Again, two fall 
exponentially toward the level of numerical truncation error, and the other 
exponentially approaches its expected value for a Kerr hole.  The quantity 
$\ell_{\rm eff}$ labeling the curves is again computed from the eigenvalue 
at late times, but the definition is now given by the 
equation $\lambda = -\ell_{\rm eff} (\ell_{\rm eff}+1)/r^2 + 2/r^2$, taken 
from the special case of Eq.~\eqref{e:GenEigProb} on a round sphere.
}
\end{figure}

\begin{figure}
\includegraphics[scale=0.35]{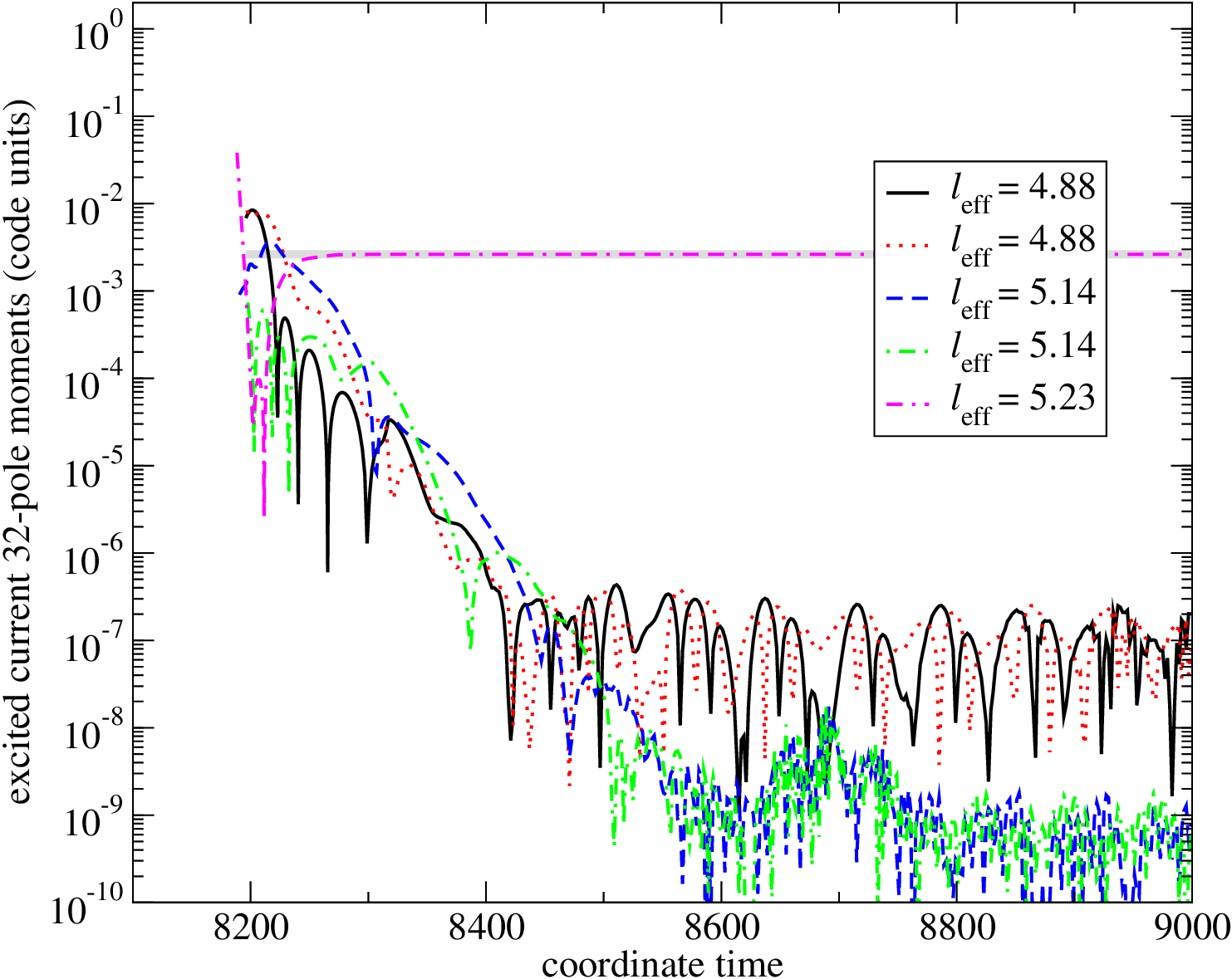}
\caption{ \label{f:triacont} Of the eleven current 32-pole moments, only five 
are allowed by the reflection symmetry to be excited.  Again, four fall 
exponentially toward the level of numerical truncation error, and the other 
exponentially approaches its expected value for a Kerr hole.  The quantity 
$\ell_{\rm eff}$ labeling the curves is defined as in Fig.~\ref{f:magoctupole}.
}
\end{figure}

\begin{figure}
\includegraphics[scale=0.35]{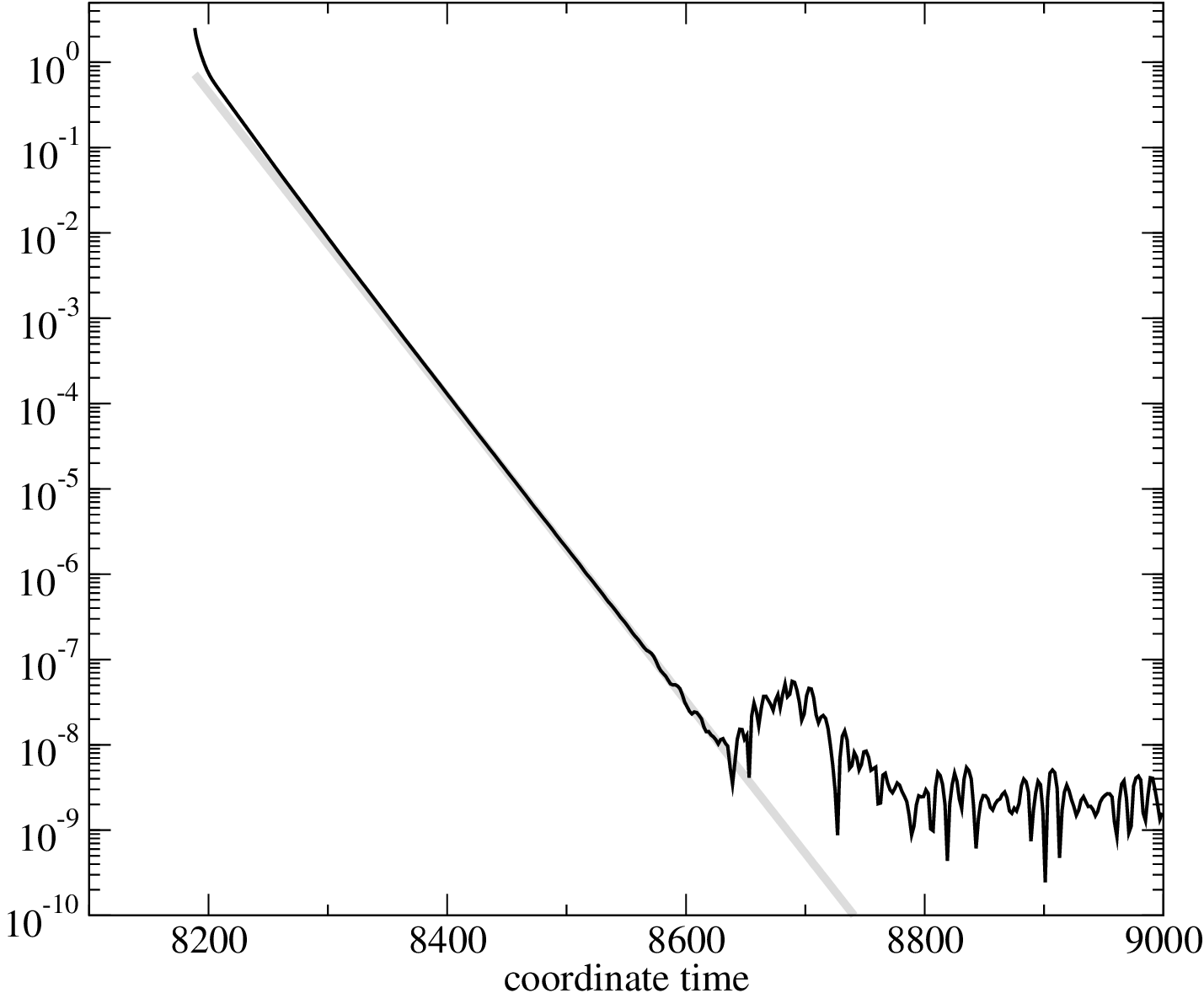}
\caption{ \label{f:KerrQuadQuadsum} The quadratic sum 
$Q(t) := \sqrt{q_1^2(t) + q_2^2(t)}$ of the two moments $q_1$ and $q_2$ that 
fall exponentially to zero in Fig.~\ref{f:quadmoments}.  The thick gray line 
shows the expected falloff of this quantity in perturbation theory.
}
\end{figure}

\begin{figure}
\includegraphics[scale=0.35]{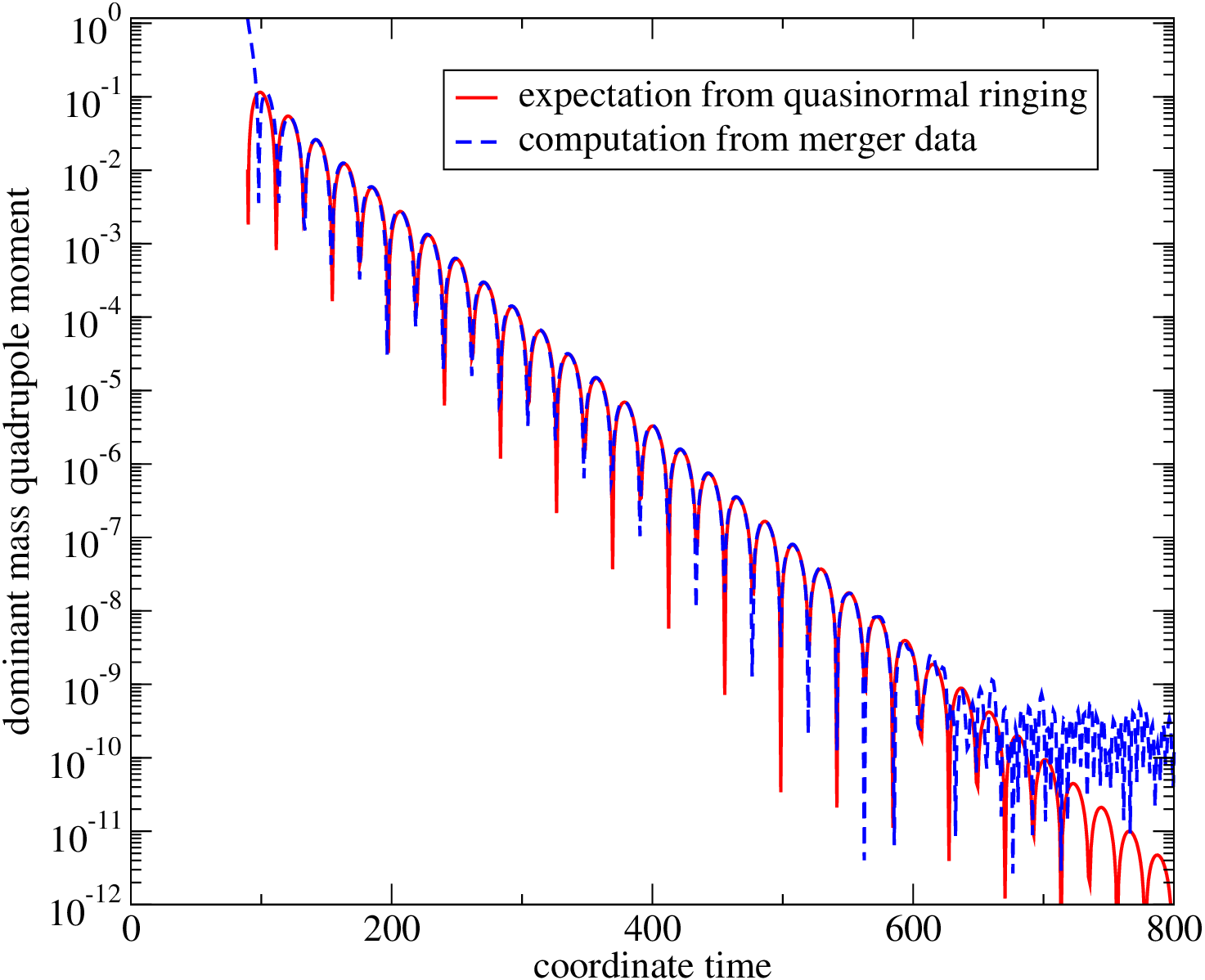}
\caption{ \label{f:SchQuadRing} Ringing of the dominant mass quadrupole moment 
as the product of a nonaxisymmetric head-on merger settles down to a 
Schwarzschild black hole.  The dashed blue curve represents numerical data, 
and the solid 
red curve is an exponentially-damped sinusoid, with frequency and damping 
corresponding to quadrupole gravitational quasinormal modes of a Schwarzschild 
black hole~\cite{Leaver1985}.  An arbitrary constant 
amplitude scaling and phase have been applied to the red curve, by eye.
}
\end{figure}

\begin{figure}
\includegraphics[scale=0.35]{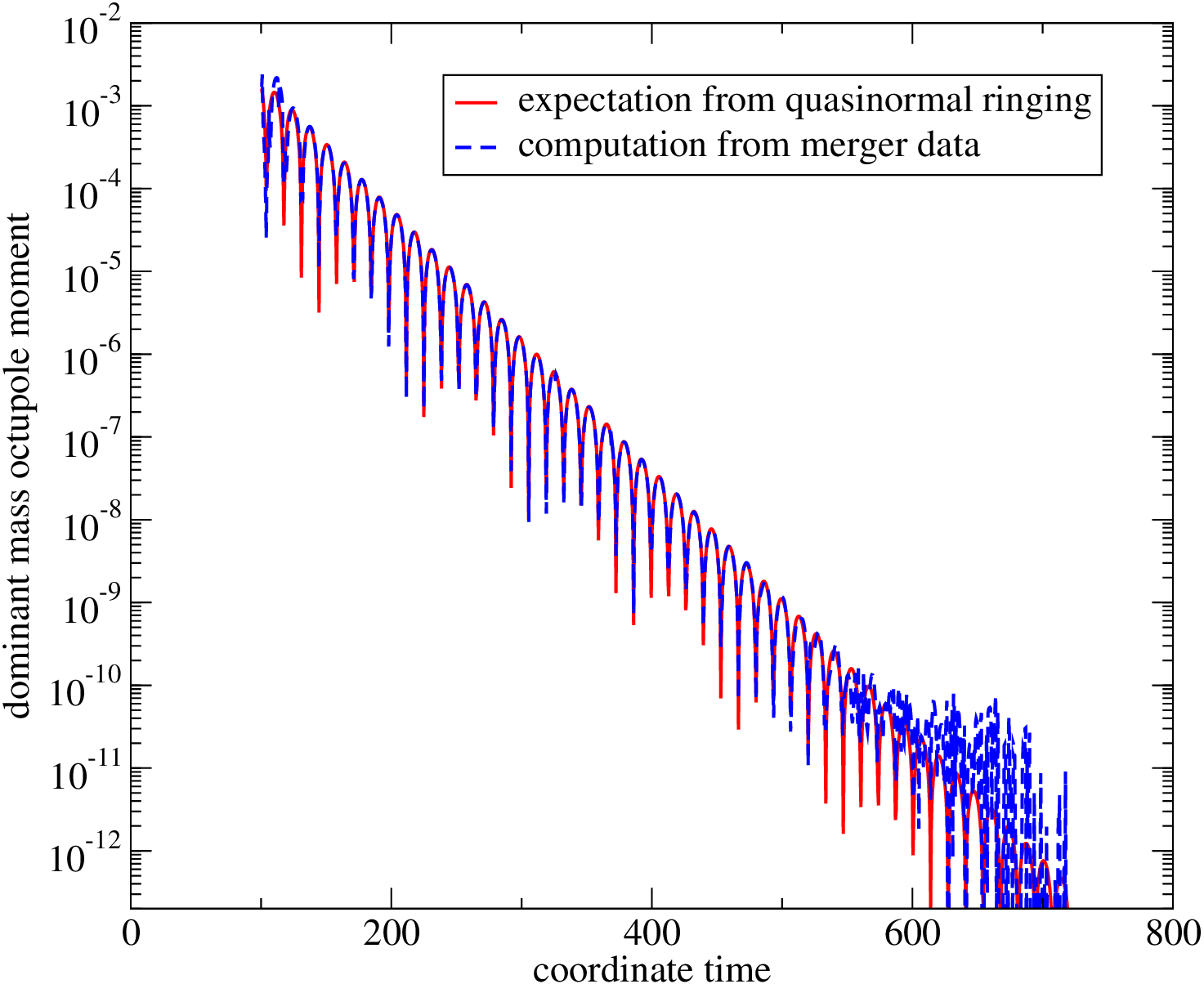}
\caption{ \label{f:SchOctRing} Ringing of the dominant mass octupole moment 
as the product of a nonaxisymmetric head-on merger settles down to a 
Schwarzschild black hole.
}
\end{figure}

\begin{figure}
\includegraphics[scale=0.35]{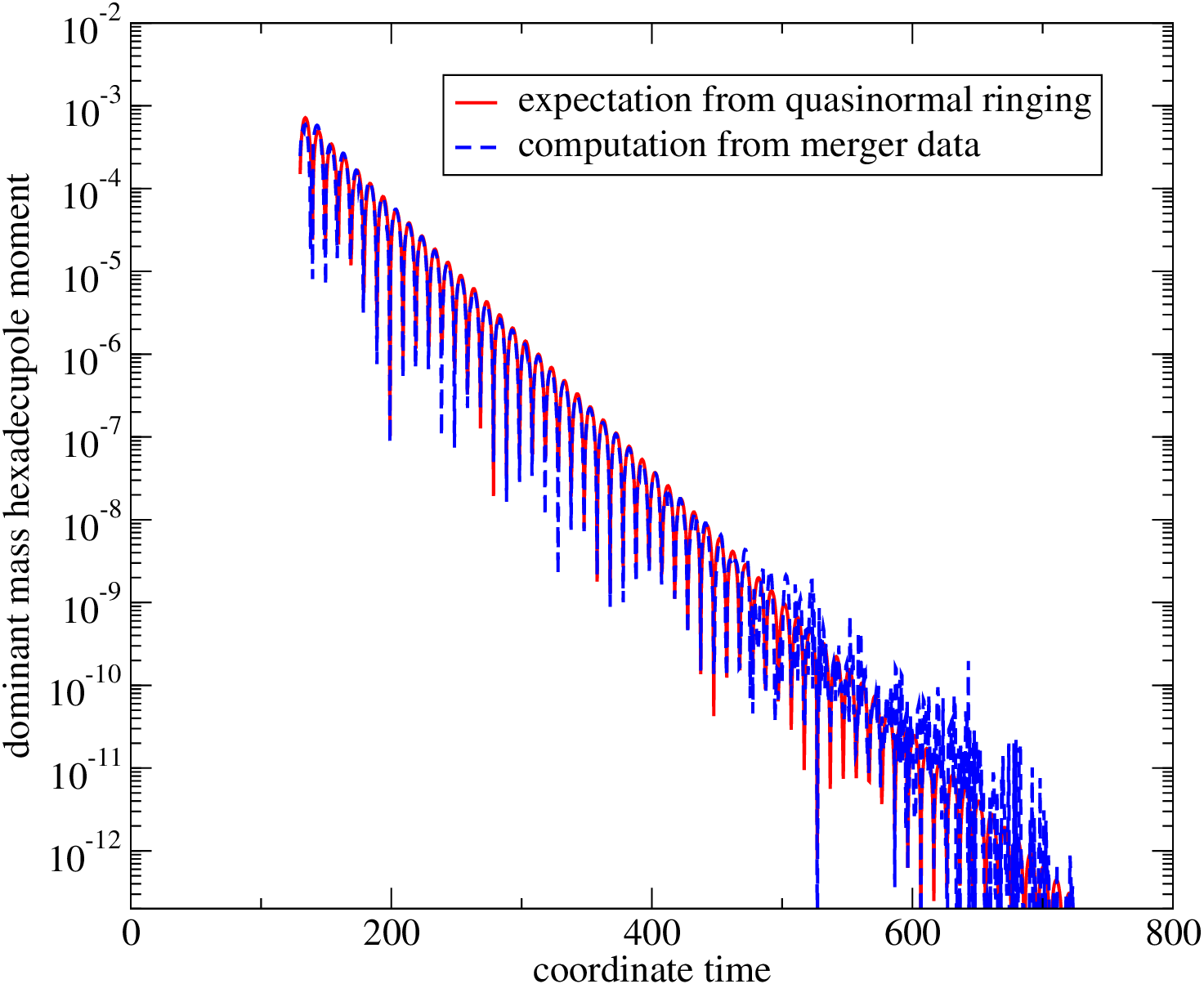}
\caption{ \label{f:SchHexRing} Ringing of the dominant mass hexadecupole moment 
as the product of a nonaxisymmetric head-on merger settles down to a 
Schwarzschild black hole.
}
\end{figure}

\begin{figure}
\includegraphics[scale=0.35]{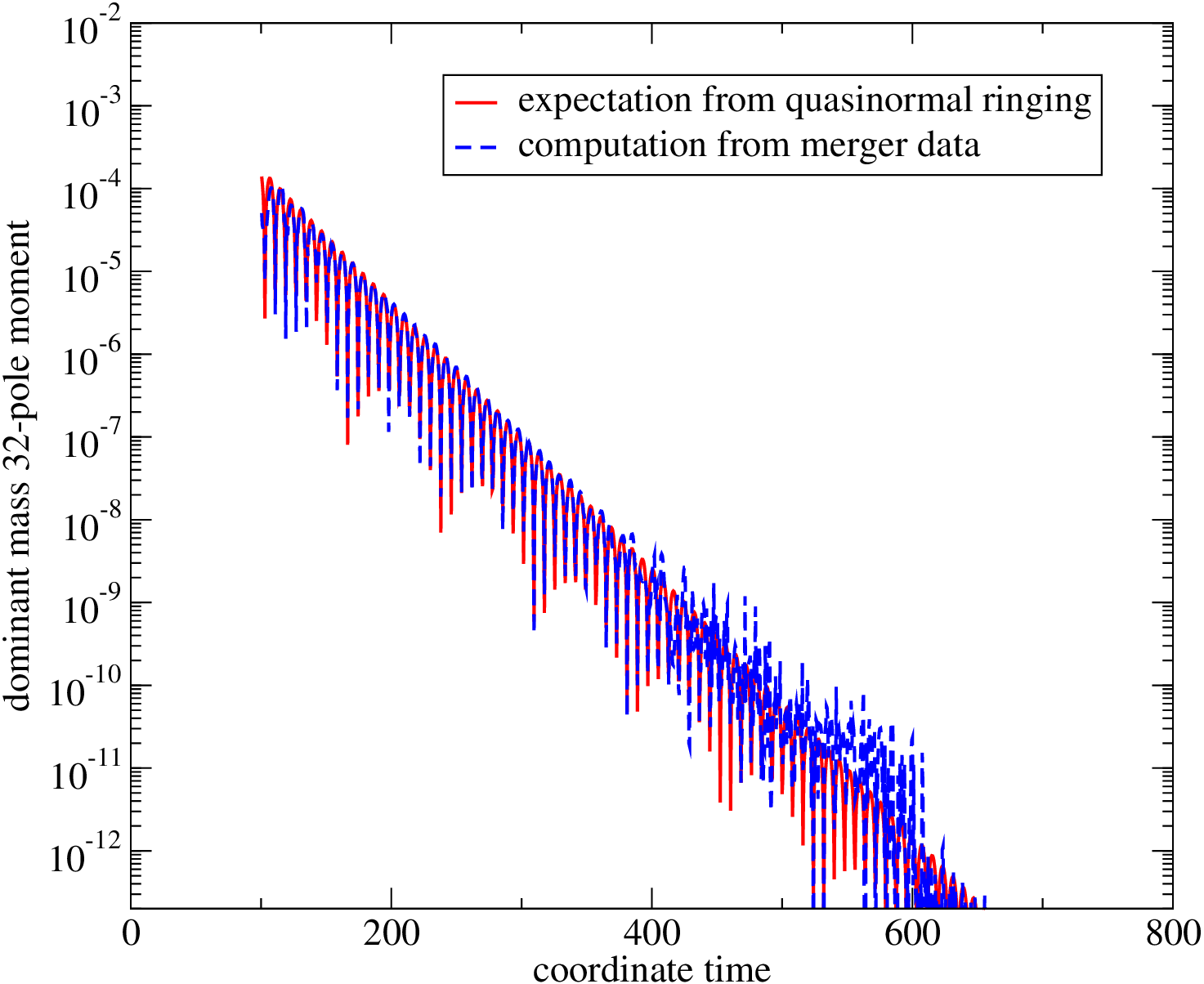}
\caption{ \label{f:SchTriRing} Ringing of the dominant mass 32-pole moment 
as the product of a nonaxisymmetric head-on merger settles down to a 
Schwarzschild black hole.
}
\end{figure}

This is an interesting tool for the study of this particular problem, because 
in the ringdown after a nonspinning black hole merger there is a  
transition from one axis of symmetry to another.  Immediately after the 
formation of a common apparent horizon, one intuitively expects this 
horizon to be ``peanut shaped,'' with an axis of approximate 
symmetry\footnote{In the case studied here, this axisymmetry would only 
be approximate, as tidal bulges would be expected to phase-shift due to 
horizon viscosity during the inspiral.  In the case of a direct head-on 
collision of nonspinning holes, this axisymmetry would be exact, and would 
be preserved even through the ringdown.}
along a line connecting the previous two individual apparent horizons.  After 
the ringdown is complete, one would expect a single black hole with symmetry 
about the axis of the initial orbital angular momentum.  This breaking and 
forming of symmetries is demonstrated 
in Fig.~\ref{f:LaplacianDipoleEigs}.  The figure presents the three 
eigenvalues of the horizon Laplacian associated with harmonics that would 
settle to the $\ell = 1$ spherical harmonics if the horizon were to become 
metrically spherical.  Two of these curves overlap at early times, a 
degeneracy due to the approximate axisymmetry of the initial ``peanut'' 
shape.  As this symmetry is broken during the ringdown, the degeneracy breaks 
and one eigenvalue eventually joins up with the third eigenvalue, 
demonstrating the eventual axisymmetry about the spin direction.\footnote{All 
figures in this paper give quantities computed in code units evaluated with 
respect to coordinate time which is also expressed in code units.  For 
context, the final black hole described in 
Figs.~\ref{f:LaplacianDipoleEigs}--\ref{f:KerrQuadQuadsum} has horizon mass 
$M \approx 1.98$ in these code units, where this mass is defined by 
the Christodoulou formula $M^2 = M_{irr}^2 + J^2/(4 M_{irr}^2)$ 
where $M_{irr} = \sqrt{A/(16\pi)}$ is the irreducible mass and $J$ is the 
quasilocal spin angular momentum defined in Appendix A of 
Ref.~\cite{Lovelace2008}.  In the simulation presented in 
Figs.~\ref{f:SchQuadRing}--\ref{f:SchTriRing}, the value of this final mass 
in code units is $M \approx 2.56$.}

With the next five eigenvalues, in Fig.~\ref{f:LaplacianQuadrupoleEigs}, 
we see the pattern again.  Again, modes are nearly degenerate at early times, 
but split off during the ringdown and reconnect as the quiescent symmetry is 
approached.  Note, however, that one degeneracy at early times is quite 
visibly broken.  This may be due, on an intuitive level, to the tidal 
interaction of the two black holes during inspiral, with shifted phase 
due to horizon 
viscosity~\cite{Thorne-Price-MacDonald, Hartle:1974, FangLovelace2005}.  
Because such tidal interaction is a 
quadrupolar effect, it would make sense for it to be less visible in the 
dipolar information of Fig.~\ref{f:LaplacianDipoleEigs}.

Degeneracies in the other eigenproblem, Eq.~\eqref{e:GenEigProb}, give a 
similar picture of the breaking and reforming of symmetries, but this problem 
gives an even more compelling picture of the relationship between symmetries 
and degeneracies.  The original motivation of Eq.~\eqref{e:GenEigProb}, as 
described in 
Appendix A of Ref.~\cite{Lovelace2008}, was to construct, in a 
sense, the closest possible approximation of an axial symmetry on a horizon 
that may not have any true symmetries at all.  One can easily show that the 
value of a given eigenvalue is proportional to the integral of the square 
of the residual in Killing's equation for the associated ``approximate Killing 
vector'' field.  Thus, when there is a true symmetry, and therefore a true 
Killing vector field, one of the eigenvalues of this problem will equal zero.  
So from plots of the eigenvalues of Eq.~\eqref{e:GenEigProb} we can see the 
breaking and forming of symmetries both indirectly, through degeneracies of 
the eigenspaces, and directly, through the value of the lowest eigenvalue.  
Figure~\ref{f:AKVOpDipoleEigs} shows the three lowest eigenvalues of this 
problem.  As noted at the end of Sec.~\ref{s:VecSphHarm}, there are no 
monopole harmonics at all for this problem, so these are the three harmonics 
that would reduce to the $\ell = 1$ harmonics if the horizon approached a 
metric sphere.  The vertical axis of the figure is now logarithmically scaled, 
to show the approach of the smallest eigenvalue to zero both at early and late 
times.  

These figures also provide a quantitative picture of the intrinsic geometry 
of the apparent horizon, and its approach at late times to the geometry of a 
slice of the Kerr horizon.  The horizontal lines in 
Figs.~\ref{f:LaplacianDipoleEigs}--\ref{f:AKVOpDipoleEigs} represent the 
expected values for these eigenvalues on a Kerr horizon of the same mass 
and spin as is measured at very late times in the simulation.  Note that this 
spin is guaranteed to be identical to the late time current dipole moment 
on the horizon (this is the main motivation for Eq.~\eqref{e:GenEigProb}), 
so agreement of the current dipole with the ``expected Kerr value'' is trivial,
however the consistency of all other multipoles, as well as these eigenspectra,
present a nontrivial demonstration that the quiescent hole is Kerr.  

Figures~\ref{f:quadmoments}--\ref{f:triacont} present the behavior of the 
multipole moments.  In Fig.~\ref{f:quadmoments}, the three excited quadrupole 
moments are shown (the other two vanish as demanded by reflection symmetry).  
One moment starts out relatively small and grows to take the value expected 
for a Kerr black hole.  The other two fall exponentially toward zero, until 
reaching the level of numerical truncation.  Figure~\ref{f:quadconvergence} 
shows the convergence of this floor of numerical error for three values of the 
resolution of the numerical simulation.  On all three simulations, the horizon 
finder and eigenvalue solver are run at the maximum 
relevant resolution, essentially the same as the angular resolution of the 
original simulations.  
Figures~\ref{f:hexadecmoments}--\ref{f:triacont} are analogous to 
Fig.~\ref{f:quadmoments}, showing higher-order multipole moments.  Again, 
all multipoles allowed by the reflection symmetry of the problem are excited 
near the moment of merger, but in each case a single moment rises to the 
expected value for a Kerr black hole of the measured final mass and spin, 
while all other multipoles decay exponentially toward zero before stopping due 
to numerical truncation.  

The fact that those multipoles that decay to zero do so exponentially raises 
the question of whether this decay can be attributed to quasinormal ringing.  
The answer to this question is clouded by a few subtleties.  For one, 
while the multipoles do appear to oscillate within an exponential envelope, 
this oscillation does not appear to be even approximately periodic, and at any 
rate occurs on a much longer timescale (compared to the exponential decay 
timescale) than the oscillations associated with quasinormal ringing.  There 
is an intuitive explanation for this.  Because the multipoles are defined with 
respect to spherical harmonics that are fixed by the intrinsic geometry of the 
horizon, changes in horizon geometry will cause changes in these harmonics.  
In particular, if the major part of the perturbation from Kerr geometry is a 
nonaxisymmetric bulge that rotates around the spin axis, then the spherical 
harmonics will be dragged along with this bulge.  Intuitively, an ideal 
``$\ell = 2$, $m = 2$'' bulge would be expected to drag the 
spherical harmonics into corotation with it, so the multipole representing 
this bulge would be expected to fall off as a pure exponential, with no 
oscillation.  In reality, the situation is more complicated, presumably due in 
part to the existence of higher multipolar structure, and in part due to the 
eventual approach to axisymmetry, causing degeneracies in the eigenproblems to 
be broken at the numerical level rather than at the analytical one.  

Properly ``unwinding'' this rotation of the harmonics would amount to a 
partial fixing of angular coordinates.  There may be sensible ways to do this, 
but we consider this somewhat oustide the scope of the current research, so 
instead we choose to ignore the oscillatory behavior, by focusing on the 
quadratic sums of multipoles associated with 
nearly-degenerate eigenspaces.  In particular, the two exponentially-decaying 
curves in Fig.~\ref{f:quadmoments} are associated with such an asymptotically 
degenerate eigenspace, and can intuitively be interpreted as real and 
imaginary parts of the ``$\ell = 2$, $m = 2$'' multipole.  Their quadratic 
sum can therefore be viewed as the overall ``magnitude'' of the quadrupolar 
part of this rotating bulge, and would be expected to fall off exponentially 
in time without oscillation.  Figure~\ref{f:KerrQuadQuadsum} shows the 
value of this quadratic sum as well as the falloff rate expected from 
perturbation theory.  We used the method due to Leaver~\cite{Leaver1985} to 
compute quasinormal frequencies in terms of the roots of two coupled complex 
continued fractions.  For the mass and spin, we used values computed at 
late times on the horizon and reported in Ref.~\cite{Scheel2008}.  

The remarkably fine agreement between this quadratic sum of multipole moments 
and the expected exponential falloff of the dominant ``$\ell = 2$, $m = 2$'' 
quasinormal mode in perturbations of the Kerr geometry makes it tempting to 
try to pick other quasinormal ringing modes out of the data.  This however  
would be a somewhat nontrivial undertaking.  For one thing, all multipoles 
defined in our formalism are computed from data directly on the horizon.  
Different radial modes of black hole perturbations would be directly 
superposed in the multipoles.  Further complicating matters in the case of a 
Kerr hole, the angular dependence of the quasinormal modes are not given by 
pure spherical harmonics, as defined here, but as solutions of the ``angular 
equation'' of the Teukolsky formalism (Eq.~(4.10) of Ref.~\cite{Teukolsky}).  Thus all multipoles would be expected to project out components of 
all radial and angular quasinormal modes, rather than cleanly projecting out 
one at a time.  What is seen in 
Fig.~\ref{f:KerrQuadQuadsum} as exponential decay is actually just the 
dominant term of a multiexponential expansion.  The problem of fitting data 
to a sum of exponentials is famously ill posed, 
so any effort to pick out higher-order ringing modes from this data would 
be quite ambitious, if possible at all.  

Many of these complications disappear if the final black hole is nonspinning.  
In that case, the quasinormal modes should have the angular dependence of pure 
spherical harmonics, so multipoles of a given order can be expected to project 
out modes of the same order (though, again, multiple radial modes would be 
expected to overlap).  Also, if there are enough degrees of reflection 
symmetry to forbid the rotation of the spherical harmonics described above, 
one might hope to recover not only the exponential falloff rates of different 
quasinormal modes, but also their frequencies of oscillation.  

Figures~\ref{f:SchQuadRing}--\ref{f:SchTriRing} demonstrate this 
recovery.  The data used here are from the ringdown after the 
collision of two black holes of nonzero antialigned spin (and therefore zero 
total angular momentum) starting from rest.  This is a simple test case 
that can be used for studying black hole kicks, and the 
particular simulation studied here will be presented in great detail for that 
goal in an upcoming paper~\cite{Lovelace2009}.  For the 
present purposes, the important points are that the final state is 
nonspinning, and that two orthogonal planes of reflection symmetry (the 
coordinate $x = 0$ and $z = 0$ planes, with the final kick being in the 
$y$ direction) hold the harmonics in place, in that they remain 
symmetric or antisymmetric under the action of the reflection symmetries.  
Figures analogous to 
Figs.~\ref{f:SchQuadRing}--\ref{f:SchTriRing} for the current multipoles show 
similar agreement, but are omitted here because they look essentially the 
same.  Incidentally, similar (though 
less detailed) agreement with quasinormal ringing 
frequencies was noted in the oscillation of the area of spatial slices of 
the event horizon in other recent simulations using the {\tt SpEC} 
code~\cite{Cohen2009}.  

One subtlety with Figs.~\ref{f:SchQuadRing}--\ref{f:SchTriRing} must be 
noted.  At late times, when these moments reach levels on the order 
of $10^{-9}$, 
the data become quite noisy.  This is a result of numerical errors, 
particularly the truncation error of the angular discretization.  However it 
appears that the numerical data in 
Figs.~\ref{f:SchOctRing}--\ref{f:SchTriRing} 
continue to decrease as the simulation goes on.  This is an artefact of the 
manner in which the data are extracted.  The numerical code computes 
essentially as many multipoles as there are grid points on the interpolated 
apparent horizon.  These multipoles must be ordered in some way.  The most 
obvious 
ordering is provided by the eigenvalues of the spherical harmonics.  However 
such an ordering is not effective when families of eigenspaces are nearly 
degenerate, as particularly in the case of ringdown to a Schwarzschild black 
hole.  To pick out particular eigenvalues in this quasinormal ringdown phase, 
we employ a simple post-processing script that chooses, at each timestep, the 
particular multipole moment that has value closest to a ``prototype'' value 
taken from the perturbation theory results shown in the red curves in these 
figures.  Simultaneously, the script checks that any chosen multipole 
corresponds to an eigenvalue which lies within a certain range, so that 
the multipole is assured to have the proper ``$\ell$'' value.  After this 
searching is carried out, we check the eigenvalue, as a function 
of time, corresponding to the chosen multipole, to ensure that it is 
smooth and therefore that the procedure has chosen a consistent multipole and 
eigenvalue dataset\footnote{Immediately after the formation of the common 
horizon, when deviations from the expectations of linearized theory are 
strongest, this script can again have trouble finding consistent multipole and 
eigenvalue datasets.  For this reason some data are omitted from the beginning 
of Figure~\ref{f:SchHexRing}, as nonsmoothness of the eigenvalues showed that 
the chosen multipoles did not represent a consistent dataset at very early 
times.}.  This procedure nicely and unambigously 
recovers physical perturbations during most of the ringdown.  However at late 
times the perturbation is small enough that the ordering ambiguity is 
particularly strong.  At late times, the script chooses the moment closest to 
the prototype value, out of the many that are oscillating quickly at small 
values due to numerical error, yet all eigenvalues in the given range are 
essentially the same, so the smoothness of the eigenvalue is no 
longer an effective tool to distinguish the correct moment from the others of 
the same $\ell$.  The matching of the numerical data to the prototype function 
is therefore given more weight than it deserves, and the data, though clearly 
flooded with numerical error, continue to fall off exponentially in time.  
Figure~\ref{f:SchQuadRing} is an exception to this behavior.  In that 
particular case, a method involving matching the spherical harmonics to 
coordinate spherical harmonics was able to unambiguously pick out the 
``correct'' harmonic.  For higher multipoles that method failed, apparently 
due to the rotation ambiguity of the coordinate spherical harmonics 
themselves. 

The quality of the agreement with standard quasinormal ringing frequencies, 
both in the approach 
to Schwarzschild geometry in Figs.~\ref{f:SchQuadRing}--\ref{f:SchTriRing} 
and in the approach to Kerr geometry in Fig.~\ref{f:KerrQuadQuadsum}, 
initially came as quite a surprise, considering that a major motivation 
for this project is a healthy skepticism for the quality of the coordinates 
in numerical simulations.  While the slicings used here are not  
arbitrary --- both simulations employ harmonic slicings during the ringdown, 
as do conventional 
treatments of black hole perturbation theory --- they are nonetheless 
{\em different} harmonic slicings than those in conventional perturbation 
theory, because the ones used in our simulations are horizon-penetrating.  
One might ask, then, how the numerical code knows to settle on a harmonic 
slicing in which these frequencies come out as expected.  The answer lies in 
the approach to stationarity.  At late times the simulations develop 
approximate stationary 
Killing vector fields $\vec \xi$, and the coordinate components of the 
spacetime metric tensor asymptote to constant values, meaning that the 
coordinates adapt themselves to the symmetry such 
that $\vec \partial_t \rightarrow \vec \xi$.  This turns the definition of 
the ringing frequencies into a geometrical statement: rather than saying 
$\partial_t^2 \Phi = -\omega^2 \Phi$, one can say 
$\xi\left(\xi\left(\Phi\right)\right) = -\omega^2 \Phi$.  In other words, 
the frequencies come out right because the coordinates adapt themselves to the 
late-term stationarity.  The process by which this adaptation occurs is, to 
our knowledge, still an open question.  

At any rate, we must caution that this recovery of standard frequencies at 
late times should by no means be taken as license to overlook gauge 
ambiguity in numerical 
simulations.  For example, it is quite tempting to associate the slight 
disagreement with perturbative results immediately after merger with nonlinear 
dynamics, however this disagreement could just as likely be due to the 
coordinates having not yet adapted to the approximate stationarity, or to 
stationarity 
simply not existing to a sufficient approximation.  All of these 
effects (and perhaps others) 
will have an impact on the ringing immediately after merger, and a detailed 
investigation of the nonlinear extension of quasinormal ringing would require 
(at least partially) slicing-invariant comparisons beyond the scope of the 
current work.  For example, one might treat the ringing of one multipole as a 
``clock'' by which to measure the frequencies of the other multipoles.

\section{Discussion}

We have presented a definition of quasilocal source multipoles on 
dynamical horizons, adapted from that in Ref.~\cite{AshtekarMultipole2004} in 
such a way 
that it can be applied to horizons without axisymmetry, while preserving 
the agreement of the current dipole moment with the spin angular momentum 
defined by Hamiltonian methods~\cite{Ashtekar2001}.  More precisely, the 
vector spherical harmonics used to project out 
current multipoles are constructed in such a way that the dipole moment 
is identical to the spin angular momentum used in Ref.~\cite{Lovelace2008}.  
The key to this generalization is the definition of spherical harmonics as 
solutions to certain eigenvalue problems on the apparent horizon.  

We have also applied this formalism to demonstrate that in a detailed and 
partially gauge-invariant sense, the binary black hole merger described 
in Ref.~\cite{Scheel2008} indeed settles to a Kerr black hole, at least in the 
neighborhood of the horizon.  There are limits to the gauge 
independence of this statement.  The work here depends heavily on the 
formalism 
of dynamical horizons~\cite{Ashtekar2003}, which are dependent on the slicing 
of spacetime (or, from a different viewpoint, are themselves invariantly 
defined yet carry unique foliations into apparent horizons that are 
compatible with the foliation of spacetime only in certain time 
slicings).  Use of a unique and  invariantly defined horizon such as the event 
horizon may be of interest (and is possible in the {\tt SpEC} 
code~\cite{Cohen2009}),
however it would not alleviate the problem of slicing dependence, as a slicing 
must be chosen at some point to break the three-dimensional horizon worldtube 
into two-dimensional surfaces on which the spherical harmonic projections are 
taken.  

A demonstration along the same lines as discussed here has been carried out 
before~\cite{Schnetter2006}, however the numerical results here are somewhat 
stronger, and our generalization of the formalism has allowed the 
consideration of a non-axisymmetric merger.  

Looking in detail at the ringdown of the multipoles, we have also recovered 
known quasinormal ringing frequencies.  The dominant exponential damping 
timescale is recovered in the ringdown to Kerr geometry, and agrees with 
results from perturbation theory.  Much more detailed results are found in 
the ringdown after a head-on collision leading to a Schwarzschild geometry, 
in which oscillation frequencies and damping timescales can be picked out 
mode by mode.

In future work we intend to study the ringdown of these 
datasets (and possibly others) on a local level, using a variant of the 
method presented in Ref.~\cite{Campanelli2008}.  

As for the multipole moments themselves, various avenues of investigation are 
open.  The methods used here could be applied to study the tidal interaction 
of black holes during fully nonlinear binary inspiral and merger, including a 
full nonlinear generalization of certain 
results~\cite{Hartle:1974, FangLovelace2005} of black hole 
perturbation theory.  As mentioned 
in Ref.~\cite{AshtekarMultipole2004}, quasilocal source multipoles might also be 
applicable in trying to find a generalization, to exact general relativity, 
of Einstein's celebrated quadrupole formula.  Related to this, one might hope 
to recover force laws at the quasilocal level, relating black hole kicks to 
products of multipoles, as is done in the asymptotic regime 
in Ref.~\cite{Schnittman2007}.  However, such an investigation would presumably 
require a satisfactory quasilocal definition of black hole linear momentum, 
which (if possible at all) appears to be beyond the realm of current 
understanding.

\acknowledgments{
I heartily thank Mark Scheel, for providing the numerical simulation 
data from Ref.~\cite{Scheel2008}, and Geoffrey Lovelace, for the data 
from Ref.~\cite{Lovelace2009}.  I also thank Saul Teukolsky 
and Larry Kidder for advice, on the project and on the manuscript.  The 
work described here was supported by NSF grants PHY-0652952, DMS-0553677, and 
PHY-0652929; NASA grant NNX09AF96G, and a grant from the Sherman Fairchild 
Foundation.
}

\bibliography{References}

\end{document}